\documentclass[%
 aip,
 amsmath,amssymb,
 reprint,%
]{revtex4-1}

\usepackage{graphicx}
\usepackage{dcolumn}
\usepackage{bm}

\usepackage[utf8]{inputenc}
\usepackage[T1]{fontenc}
\usepackage{mathptmx}
\usepackage{etoolbox}

\usepackage{bm}
\usepackage{amsthm} 

\usepackage[english]{babel}
\makeatletter
\def\@email#1#2{%
 \endgroup
 \patchcmd{\titleblock@produce}
  {\frontmatter@RRAPformat}
  {\frontmatter@RRAPformat{\produce@RRAP{*#1\href{mailto:#2}{#2}}}\frontmatter@RRAPformat}
  {}{}
}%
\makeatother
\begin{document}

\preprint{AIP/123-QED}

\title[Diffeomorphism Invariance and Background Independence]{Diffeomorphism Invariance and Background Independence}
\author{K. Tjandra}
 \email{ktj@nus.edu.sg}
\affiliation{National University of Singapore
}%
\author{K. Singh}%
\affiliation{National University of Singapore
}%

\date{\today}

\begin{abstract}
This paper answers examines the relationship between Diffeomorphism Invariance and Background Independence. First, a review of the relationship between Background Independence, General Relativity (GR) and pre-GR theories are given. Then, a wide range of other conceptions of background independence is discussed. It is shown that the definition of Background Independence is fluid and can mean different things to different philosophers and/or physicists. Most pertinently, the paper addresses the question of what kind of background independence is implied by a mathematical criterion of diffeomorphism invariance or in what sense is diffeomorphism invariance background independence. Lastly, the concept of haecceity in relation to diffeomorphism invariance is discussed.
\end{abstract}

\maketitle

\begin{quotation}

\end{quotation}

\section{\label{sec:intro}Introduction}

Why would Diffeomorphism Invariance be desirable for a theory? The answer to this profound question is often discussed in the context of the closely related concept of Background Independence. Whereas Diffeomorphism Invariance is a mathematical characteristic of a particular variable or theory, and therefore is practically analyzable as has been demonstrated in this thesis, Background Independence is largely interpretational or philosophical in nature. As such, the relationship between Background Independence and Diffeomorphism Invariance or other properly mathematical concepts such as the absolute object, is not immediately apparent. To be clear, any notion of Background Independence need an anchor in the form of a clear mathematical condition, but it is not evidently obvious what that clear mathematical condition ought to be. 

A range of answers have been provided with regards to the significance of Diffeomorphism Invariance and Background Independence in the wider literature. On one extreme of the spectrum, a largely philosophical argument by Teitel in \cite{Teitel} argues that Diffeomorphism Invariance and Background Independence is not a non-empirical virtue. Teitel claims that given two theories with the same empirical predictive power (two theories that can explain the same empirical data), one theory with Diffeomorphism Invariance and one theory without Diffeomorphism Invariance, we would not be justified in adjudicating between these two theories on the basis of their Diffeomorphism Invariance or Background Independence. Given one theory with Diffeomorphism Invariance and one theory without, Teitel would simply ask: which one has the better empirical predictive power? In the middle of the spectrum, Physicists such as Smolin in \cite{Smolin}, describe Diffeomorphism Invariance and Background Independence as merely part of a strategy, a relational strategy. Smolin describes how the strategy of producing more and more relational theories has manifestly produced empirically better and better theories. This fact would then justify the Physicist' preference for more relational theories, in particular, theories with Diffeomorphism Invariance and Background Independence. On the opposite extreme of the spectrum, one may claim, as Stachel did in \cite{Stachel1} and \cite{Stachel2}, that Diffeomorphism Invariance is an unalterable discovered fact of nature that theories following the advent of GR that is written in the language of manifolds ought to have. No new theories should contradict this requirement as it is inconceivable how such a theory would be approximated by GR in its appropriate limits. 

This paper shall contribute in this discourse regarding Diffeomorphism Invariance and Background Independence, first by discussing Background Independence in the original context of the advent of GR as discussed by Pooley in \cite{Pooley}, second by discussing Background Independence in relation to the concept of haecceity as discussed by Stachel in \cite{Stachel1}.

\section{Background Independence, GR, and Pre-GR Theories}

The advent of Einstein's GR theory brought about a significant paradigm shift in Physics. A totally new way of conceiving space, described with the new mathematics of differential geometry, was suddenly found to be foundational. Space that was largely functioning as a container for matter and energy previously was suddenly dynamical. In this profoundly different theory, the Physicist is led to question what is the most fundamental difference that differentiates GR from all theories of Physics prior to it. Beyond the newness of the mathematics that was used to formulate GR, is there a more fundamental philosophical implication that sprung out of GR?

As Pooley pointed out in \cite{Pooley}, Background Independence is conceived to be one fundamental philosophical implication of GR that differentiates it from pre-GR theories. Background Independence as it was first conceived, is an intuitive idea. In some intuitive way, it was apparent that GR fundamentally relies less on the background structures than its predecessors such as Special Relativity (SR), the immediate precursor to GR. These background structures may include the absolute space and time of Newtonian Mechanics, the Minkowski Metric of SR, etc. However, beyond the specific structures of these theories, is there an overarching mathematical requirement that differentiates GR from pre-GR theories and therefore renders a theory Background Independent or not Background Independent.  

Diffeomorphism Invariance to some seem to be the specific mathematical requirement that renders a theory Background Independent. After all, while GR possesses Diffeomorphism Invariance, SR does not. SR's Minkowski's Metric, is not invariant under a general diffeomorphism. If Diffeomorphism Invariance is indeed Background Independence, this thesis have shown AGT and TEGR to be Background Independent. However, Pooley in \cite{Pooley} pointed out that one can actually reformulate SR to be Diffeomorphism Invariant. Pooley shows that if SR instead is formulated using the mathematical structure of GR with space left to be vacuum, without any sources of gravity that affects curvature, that reformulation of SR will be Diffeomorphism Invariant. This SR reformulated as GR with flat connection and constant metric, that locally is equivalent to the Minkowski Metric, is empirically equivalent to SR, and yet is as Diffeomorphism Invariant as GR. Thus, the intuitively Background Dependent theory can have Diffeomorphism Invariance when written in the appropriate formulation.  

Therefore, if one takes Background Independence to necessarily be the differentiating characteristic that separates GR from pre-GR theories, on account of Pooley's Background Independent formulation of SR, Diffeomorphism Invariance cannot simply be equated with Background Independence. However, it would proof useful to remember that the idea of Background Independence itself does not have a universal definition. Considering the GR versus pre-GR differentiating ability of Background Independence to be definitional is indeed unique to Pooley.   

\section{Other Conceptions of Background Independence}

As Pooley simply rejected the identification between Diffeomorphism Invariance and Background Independence, he offered another way of relating Diffeomorphism Invariance and Background Independence. He pointed out that while SR has both Diffeomorphism Invariant and non Diffeomorphism Invariant formulation, GR can only be formulated in a Diffeomorphism Invariant manner. There is no GR without Diffeomorphism Invariance. As such, one may define that a theory is Background Independent if in all formulations, the theory is necessarily Diffeomorphism Invariant. This definition of Background Independence is clearly a stronger requirement than merely Diffeomorphism Invariance.Why would Diffeomorphism Invariance be desirable for a theory in the first place? The answer to this profound question is often discussed in the context of the closely related concept of Background Independence. Whereas Diffeomorphism Invariance is a mathematical characteristic of a particular variable or theory, and therefore is practically analyzable as has been demonstrated in this thesis, Background Independence is largely interpretational or philosophical in nature. As such, the relationship between Background Independence and Diffeomorphism Invariance or other properly mathematical concepts such as the absolute object, is not immediately apparent. To be clear, any notion of Background Independence need an anchor in the form of a clear mathematical condition, but it is not evidently obvious what that clear mathematical condition ought to be. 

A range of answers have been provided with regards to the significance of Diffeomorphism Invariance and Background Independence in the wider literature. On one extreme of the spectrum, a largely philosophical argument by Teitel in \cite{Teitel} argues that Diffeomorphism Invariance and Background Independence is not a non-empirical virtue. Teitel claims that given two theories with the same empirical predictive power (two theories that can explain the same empirical data), one theory with Diffeomorphism Invariance and one theory without Diffeomorphism Invariance, we would not be justified in adjudicating between these two theories on the basis of their Diffeomorphism Invariance or Background Independence. Given one theory with Diffeomorphism Invariance and one theory without, Teitel would simply ask: which one has the better empirical predictive power? In the middle of the spectrum, Physicists such as Smolin in \cite{Smolin}, describe Diffeomorphism Invariance and Background Independence as merely part of a strategy, a relational strategy. Smolin describes how the strategy of producing more and more relational theories has manifestly produced empirically better and better theories. This fact would then justify the Physicist' preference for more relational theories, in particular, theories with Diffeomorphism Invariance and Background Independence. On the opposite extreme of the spectrum, one may claim, as Stachel did in \cite{Stachel1} and \cite{Stachel2}, that Diffeomorphism Invariance is an unalterable discovered fact of nature that theories following the advent of GR that is written in the language of manifolds ought to have. No new theories should contradict this requirement as it is inconceivable how such a theory would be approximated by GR in its appropriate limits. 

In this chapter, this thesis shall contribute in this discourse regarding Diffeomorphism Invariance and Background Independence in the context of AGT and to a lesser extent TEGR, first by discussing Background Independence in the original context of the advent of GR as discussed by Pooley in \cite{Pooley}, second by discussing Background Independence in relation to the concept of haecceity as discussed by Stachel in \cite{Stachel1}, and lastly by explicating what Diffeomorphism Invariance as Background Independence descriptively mean for AGT or other theories. 

\section{Diffeomorphism Invariance as Background Independence}

If Diffeomorphism Invariance cannot be the differentiating characteristic of General Relativity as Pooley argued, does that mean that Diffeomorphism is definitely not, or even not related to, Background Independence? This question is meaningless. As was pointed out earlier, the concept of Background Independence itself cannot be unambiguously defined. Rather than taking a vague intuitive idea of Background Independence as the starting point, what if Diffeomorphism Invariance is taken to be the starting point of the conversation regarding Background Independence. In other words, we will ask the question of what philosophical or interpretational insight can we glean from Diffeomorphism Invariance and then recognize that insight as an aspect of Background Independence. 

Diffeomorphism Invariance implies that the spacetime manifold that underlies a theory is not a given. Through diffeomorphism, a manifold can be said to be mapped into another manifold that is isomorphic to that manifold. (Another way of describing this idea is that of the shape of the manifold. However, the idea of the shape of the manifold is vague since a manifold is endowed with "shape" only when connections are defined beforehand.) As such, Diffeomorphism Invariance means that instead of one fixed manifold that is a given for a theory, a whole equivalence class of manifolds can underlie the theory. This manifold can be described as a background, and a Diffeomorphism Invariant theory is independent of such background. In the sense of the previous paragraph, AGT is independent of the manifold ("shape") background, so is TEGR, and so is GR. 

Smolin in \cite{Smolin} notes that GR is only a partly relational theory. We might say that GR is only a partly Background Independent theory. As Smolin pointed out, there are background structures that are fixed in GR: dimension, topology, differential structure, etc. These are indeed background structures. Therefore, not even GR can be described to be completely Background Independent. Rather, there are different degrees of Background Independence. Diffeomorphism Invariance is one such degree of Background Independence. A theory that has Diffeomorphism Invariance, such as AGT, TEGR, and GR, is more Background Independent than a theory without, since the manifold ("shape") is not fixed. 

\section{Diffeomorphism Invariance and Haecceity}

One last philosophical idea relating to Background Independence and Diffeomorphism Invariance will be discussed in this section, that of the lack of haecceity. One may be tempted to describe the lack of haecceity as the sense in which Diffeomorphism Invariance can be described as a degree of Background Independence. However, the author argues that such a conclusion is unfounded. 

Haecceity comes from the Latin word \textit{haec} which means, this. As such, haecceity literally means this-ness. Haecceity refers to our ability to individuate entities. It needs to be clarified that when the Physicist talks about haecceity as in \cite{Stachel1} \cite{Stachel2}, what we mean is qualitative haecceity as opposed to non-qualitative haecceity. For the purpose of the arguments in this thesis, it is suffice to say that we are concerned with whether, under a diffeomorphism, a theory remains empirically equivalent, and/or empirically valid. Furthermore, what is empirical is necessarily qualitative even if the idea of qualitative properties itself is more complex than merely the empirical non-empirical distinction. For a complete discussion of qualitative and non-qualitative properties, refer to \cite{Cowling}. As such, when we refer to haecceity, we mean whether entities can be individuated through empirical means. In particular, we are concerned with whether spacetime points have haecceity. 

Stachel in \cite{Stachel1}\cite{Stachel2} argues that if we require the fundamental entities of our theories to have no haecceity, then one ought to be able to permutate all the constituent fundamental entities of the theory, with no empirical consequences. He refers to this principle as the Principle of Maximal Permutability. He then argues that if the manifold is a constituent entity of a theory then this principle takes the form of Diffeomorphism Invariance, as diffeomorphisms allow points to be mapped into other points in a continuous manner since the constituent entity of the manifold is continuous. 

We shall raise two objections to Stachel's argument. The first being that since Newtonian Mechanics, the space on which the theory is written already lack hacceity, even when general diffeomorphisms are disallowed in the theory. As we know, we can do spatial translations and Newtonian laws of motion or gravity will remain valid. These spatial translations, or even temporal translations, does map space and time points to different space and time points, and no one mapping is disallowed. However, this class of transformations cannot be extended into general diffeomorphisms because diffeomorphisms can break relations between the constituent spacetime points, namely the relation of lengths. General diffeomorphisms are disallowed in Newtonian Mechanics not because individual space and time points have haecceity and therefore cannot be mapped or "permutated".

The second objection is that diffeomorphisms are not in the sense of Stachel's arguments, maximal. Diffeomorphisms are not free permutations of the spacetime points. There are still relations within the manifold that needs to be preserved in diffeomorphisms, namely the continuity and smoothness. As such, the idea of Maximal Permutability itself is not representative of Diffeomorphism Invariance. Just as lengths need to be preserved in Newtonian Mechanics, continuity and smoothness needs to be preserved in in Diffeomorphism Invariant theories, such as AGT, TEGR, and GR. The group of Diffeomorphism is larger than the subgroup of translations. In this sense, Diffeomorphism Invariant theories are more Background Independent as the background relational structures are less compared to non Diffeomorphism Invariant theories, but in no way are Diffeomorphism Invariant theories lack haecceity more than Translational Invariant theories. 

Diffeomorphism Invariant theories indeed allow spacetime points to be mapped into spacetime points, and no one specific "permutation" or mapping is disallowed. Therefore, Diffeomorphism Invariance implies the lack of haecceity, but they are not equivalent. In fact, Diffeomorphism Invariance is a stronger condition than the lack of haecceity.

\bibliography{aipsamp}

\end{document}